\documentclass[12pt,preprint]{aastex}

\begin{document}

\title{Neutral Hydrogen Absorption Toward XTE~J1810--197: the
Distance to a Radio-Emitting Magnetar}

\author{
  Anthony H. Minter,\altaffilmark{1}
  Fernando Camilo,\altaffilmark{2}
  Scott M. Ransom,\altaffilmark{3}
  Jules P. Halpern,\altaffilmark{2}
  and Neil Zimmerman\altaffilmark{2} }


\altaffiltext{1}{National Radio Astronomy Observatory, Green Bank,
WV 24944.}
\altaffiltext{2}{Columbia Astrophysics Laboratory, Columbia University,
New York, NY 10027.}
\altaffiltext{3}{National Radio Astronomy Observatory, Charlottesville,
VA 22903.}

\def\magnetar{XTE~J1810--197}
\newcommand{\sgr}{\object{SGR~1806--20}}
\newcommand{\snr}{\object{G11.2--0.3}}
\newcommand{\darkcloud}{\object{G10.74--0.13}}
\newcommand{\wthirtyone}{\object{W~31}}
\newcommand{\gtentwo}{\object{G10.2--0.3}}
\newcommand{\gtenthree}{\object{G10.3--0.1}}
\newcommand{\gtensix}{\object{G10.6--0.4}}
\newcommand{\gtenzero}{\object{G10.0--0.3}}
\newcommand{\lbv}{\object{LBV~1806--20}}
\newcommand{\ourdistance}{3.1--4.3\,kpc}
\newcommand{\flatdist}{$3.4_{-0.7}^{+0.5}$\,kpc}
\newcommand{\weinerdist}{$4.0_{-0.8}^{+0.3}$\,kpc}
\newcommand{\englemaierdist}{$3.7\pm0.6$\,kpc}
\newcommand{\englemaierdistboth}{$3.3_{-0.2}^{+0.2}$\,kpc or $4.0_{-0.4}^{+0.3}$\,kpc}

\newcommand{\nraoblurb}{The National Radio Astronomy Observatory is a
facility of the National Science Foundation operated under cooperative
agreement by Associated Universities, Inc.}

\begin{abstract}
We have used the Green Bank Telescope to measure \ion{H}{1} absorption
against the anomalous X-ray pulsar \magnetar. Assuming a flat rotation
curve, we find that \magnetar\ is located at a distance of \flatdist.
For a rotation curve that incorporates a model of the Galactic bar,
we obtain a distance of \weinerdist.  Using a rotation
curve that incorporates a
model of the Galactic bar and the spiral arms of the Galaxy, the distance
is \englemaierdist.  These values are consistent with the distance
to \magnetar\ of about 3.3\,kpc derived from its dispersion measure,
and estimates of 2--5\,kpc obtained from fits to its X-ray spectra.
Overall, we determine that \magnetar\ is located at a distance of
$3.5\pm0.5$\,kpc, possibly not far in front of the infrared dark cloud
\darkcloud.  We also used the GBT in an attempt to measure absorption
in the OH ${\rm ^2\pi_{3/2}(J=3/2)}$ lines against \magnetar.  We were
unsuccessful in this, mainly because of its declining radio flux density.
Analysis of \ion{H}{1} 21\,cm, OH ${\rm ^2\pi_{3/2}(J=3/2)}$, and
$^{12}$CO$(2\rightarrow 1)$ emission toward \magnetar\ allows us to
place a lower limit of $N_{\rm H} \ga 4.6 \times 10^{21}$ cm$^{-2}$ on
the non-ionized hydrogen column density to \magnetar, consistent with
estimates obtained from fits to its X-ray spectra.

\end{abstract}

\keywords{ISM: clouds --- ISM: individual (G10.74--0.13) --- pulsars:
individual (XTE~J1810--197) --- radio lines: ISM }

\section{Introduction}

Prior to the detection of pulsed radio emission from the anomalous
X-ray pulsar (AXP) \magnetar\ by \citet{camilo}, pulsed emission
from the dozen known magnetars had been detected in X-rays in all
instances, and in one case at optical wavelengths.  Estimating the
distance to a magnetar has relied on associating it with a supernova
remnant (SNR) of known distance, or by fitting its X-ray spectrum and
parameterizing the energy-dependent absorption by interstellar gas
along the line of sight with a non-ionized hydrogen column density
$N_{\rm H}$ \citep[][]{mccammon}.  Using standard relations, $N_{\rm
H}$ is related to visual extinction $A_V$, which is turned into a
distance estimate using mean values in the Galactic plane \citep[see,
e.g., \S6 of][]{firstdist}, or is directly calibrated as a function of
distance using stars of known luminosity in the field \citep{redclump}.
Sometimes, the probable location of the X-ray source in a well-studied
star cluster can be used to infer its distance \citep{muno}.

The unique detection of pulsed radio emission from \magnetar\ allows
one to determine its distance using methods that are not applicable to
other magnetars.  The dispersion measure (DM, the total column density
of free electrons along the line of sight) was obtained upon discovery
of the radio pulsations \citep{camilo}.  A model for the Galactic free
electron distribution then yields a distance estimate, in this case $d
\approx 3.3$\,kpc using the most recent model \citep{ne2001}.  Also,
bright, pulsed radio emission allows kinematic distance limits to be
obtained by observing spectral lines that are seen in absorption against
the magnetar, as we report here using \ion{H}{1}.

Obtaining a reliable distance to \magnetar\ allows for a precise
determination of the luminosity of the star based on its measured flux
in a variety of wavebands.  The distance also allows a proper motion
\citep[see][]{helfand} to be converted into a tangential velocity.
Magnetars are thought to be
very young neutron stars, and are expected to be found near star
forming regions and/or spiral arms.  Knowing the distance to \magnetar\
allows this prediction to be tested in this case.  Besides providing a
kinematic distance, the \ion{H}{1} absorption spectrum can also give an
independent estimate of $N_{\rm H}$, which may be compared to results
from X-ray spectral fitting.

In \S~2 we present the \ion{H}{1} and OH observations and data analysis.
This is followed by a determination of the hydrogen absorption spectra
toward \magnetar, in \S~3, and of its kinematic distance in \S~4.
In \S~5 we comment briefly on some features of the neutral hydrogen
toward \magnetar, and in \S~6 on the OH absorption limits.  We obtain a
limit on the hydrogen column density in \S~7, and comment on models of
the line of sight toward \magnetar\ in \S~8.  We conclude in \S~9 with
a discussion of our main results.

\section{Observations and Data Analysis}

\subsection{\ion{H}{1} 21\,cm Absorption Observations}

\magnetar\ was observed with the National Radio Astronomy Observatory
(NRAO\footnote{\nraoblurb}) Robert C. Byrd Green Bank Telescope (GBT)
for approximately 3 hr on 2006 June 6 in order to measure the \ion{H}{1}
21\,cm line absorption.  The GBT has an unblocked aperture, a spatial
resolution of $9\farcm2$ at 21\,cm, and a system temperature on cold
sky of 18\,K.  The NRAO spectral processor (SP), an FFT spectrometer,
was used as the detector.  The SP provides good dynamic range for the
observations via its 32-bit sampling, and was used in its pulsar mode
whereby it accumulates spectra that are folded synchronously with the
pulsar period.  The SP was configured to have 1024 channels for each
linear polarization across a bandwidth of 2.5\,MHz, producing a spectral
resolution of 0.52\,km\,s$^{-1}$ per channel.  We recorded 128 spectra
(phase bins) evenly spaced in time across each individual period of
\magnetar.  The duty cycle of this 5.54\,s pulsar was such that 4--5 of
these bins contained pulsed flux.

\subsection{OH ${\rm\bf ^2\pi_{3/2}(J=3/2)}$ Absorption Observations}

\magnetar\ was observed with the GBT for approximately 38\,hr in order
to measure absorption in the 1612, 1665, 1667 and 1720\,MHz ${\rm
^2\pi_{3/2}(J=3/2)}$ ground state transitions of OH.  Observing details
are given in Table~\ref{table:absobs}.  The GBT's angular resolution
is approximately $8'$ for each of the OH transitions.  The spectral
processor was also used for these measurements and was configured to
provide 256 spectral channels for each linear polarization across each
of the four OH transitions.  A bandwidth of 0.625\,MHz was used, giving
a spectral resolution of 0.44\,km\,s$^{-1}$ per channel.  We chose this
narrower bandwidth (with higher frequency resolution) since we already
knew which velocities displayed \ion{H}{1} absorption.  As with the
\ion{H}{1} measurements, the pulsed flux of \magnetar\ was detected in
4--5 phase bins.

The GBT auto-correlation spectrometer (ACS) was used to obtain pulsar
``off'' spectra for the four OH lines.  The data were obtained in six
position-switching observations, each consisting of two minutes on source
followed by two minutes off source, resulting in an effective integration
time of 12 minutes.  An off position two minutes of time in R.A. offset
from the position of \magnetar\ was used so that approximately the same
hour-angle coverage was obtained for the on and off positions.  The ACS
was configured to observe all four of the OH ${\rm ^2\pi_{3/2}(J=3/2)}$
lines simultaneously with 8192 channels within a bandwidth of 12.5\,MHz,
resulting in a spectral resolution of 0.27\,km\,s$^{-1}$.
These observations allow for a better calibration of the pulsar ``off''
OH spectra.

\subsection{Data Reduction}

The data were analyzed using a method similar to that described
in \cite{weisberg} and \cite{minter}.  For each individual pulse
of \magnetar\ and for each spectral line we measured the flux
$T(\nu,\phi,\mathcal{P})$ in terms of the equivalent brightness
temperature $T$, versus frequency $\nu$, phase $\phi$ across the
pulse (equivalent to time), and polarization $\mathcal{P}$.  For each
polarization and pulse, the data were divided into two separate parts,
a pulsar ``on'' spectrum and a pulsar ``off'' spectrum:
\begin{eqnarray}
{T_{\rm on}(\nu,\mathcal{P})} & = & {{\sum_{\phi_{\rm on}} \left( T(\nu,\phi_{\rm on},\mathcal{P})
\left[ { \left< T(\nu, \phi_{\rm on}, \mathcal{P})\right>_{\nu({\rm noabs})} \over T_{\rm sys} } \right]^2 \right)
\over
\sum_{\phi_{\rm on}} \left[ { \left< T(\nu, \phi_{\rm on}, \mathcal{P}) \right>_{\nu(\rm noabs)} 
\over T_{\rm sys}} 
\right]^2} } \\
& & \nonumber \\
& & \nonumber \\
{T_{\rm off}(\nu,\mathcal{P})} & = & { { \sum_{\phi_{\rm off}} 
T(\nu,\phi_{\rm off},\mathcal{P}) \over N_{\phi_{\rm off}}} } 
\end{eqnarray}
where ${\phi_{\rm on}}$ are the phase bins when the pulsar is on,
${\phi_{\rm off}}$ are the phase bins when the pulsar is off,
${N_{\phi_{\rm off}}}$ is the total number of phase bins when the
pulsar is off and ${\rm \left< \right>_{\nu({\rm noabs})}}$ denotes
averaging over the frequencies that do not show absorption in the final
spectrum\footnote{An iterative approach in the data reduction is necessary
in order to determine ${\rm \nu(noabs)}$.}.  A pseudo-absorption spectrum
for the $i$th pulse is then formed by taking the difference between the
pulsar on and pulsar off spectra,
\begin{displaymath}
T_{\rm on}^i({\nu}, \mathcal{P}) - T_{\rm off}^i({\nu}, \mathcal{P}) 
 = 
\left( T_{\rm p}^i(\nu, \mathcal{P})e^{-\tau(\nu)} + T_{\rm H~I}({\nu}) + T_{\rm sys} 
\right) - \left( T_{\rm H~I}({\nu}) + T_{\rm sys} \right)
\end{displaymath}
\begin{equation}
 =  
T_{\rm p}^i(\nu, \mathcal{P}) e^{-\tau(\nu)} 
\end{equation}
where ${T_{\rm p}}$ is the brightness temperature of the pulsar.  We then
take a weighted average of the pseudo-absorption spectrum
\begin{equation}
{\left< T_{\rm p}(\nu, \mathcal{P}) \right> e^{-\tau(\nu)} = 
{ \sum_i \left( T_{\rm p}^i(\nu, \mathcal{P}) e^{-\tau(\nu)} 
\left[ { \left< T_{\rm p}^i(\nu, \mathcal{P}) \right>_\nu({\rm noabs}) \over T_{\rm sys} } 
\right]^2 \right)
\over \sum_i \left[ { \left< T_{\rm p}^i(\nu, \mathcal{P}) \right>_\nu({\rm noabs}) \over 
T_{\rm sys} } \right]^2 }} \label{eq:wght}
\end{equation}
and store the weights for later use when the polarizations are averaged
together.  This weighting is proportional to the signal-to-noise
ratio obtained for each pulse of \magnetar.  A third-order orthogonal
polynomial was fitted to the $\left<T_{\rm p}(\nu, \mathcal{P})\right>
e^{-\tau(\nu)}$ spectrum in order to determine the intrinsic pulsar
brightness, $T_{\rm p}^{\rm fit}(\nu, \mathcal{P})$, at all frequencies,
i.e., by extrapolating $\left<T_{\rm p}(\nu, \mathcal{P})\right>$ across
the absorption features.  This yielded the absorption spectrum
\begin{equation}
e^{-\tau({\nu})} = 
{ \left< T_{\rm p}(\nu, \mathcal{P}) \right> e^{-\tau(\nu)} \over
T_{\rm p}^{\rm fit}({\nu, \mathcal{P}}) }
\end{equation}
for each polarization.  The absorption spectra for the
different polarizations were then averaged using the weights from
equation~(\ref{eq:wght}) to create the final, measured absorption spectrum
for each spectral line.

For the \ion{H}{1} data, the pulsar off spectrum (i.e., the normal
\ion{H}{1} emission spectrum) was converted from detector counts to a
Kelvin scale using a calibrated noise diode that was injected during
two-minute calibration scans.  Observations of the IAU \ion{H}{1}
standard source S6 were also made to put the pulsar off spectrum on the
IAU standard brightness temperature scale.

\section{The Absorption Spectra Toward XTE~J1810--197}

\begin{figure}
\plotone{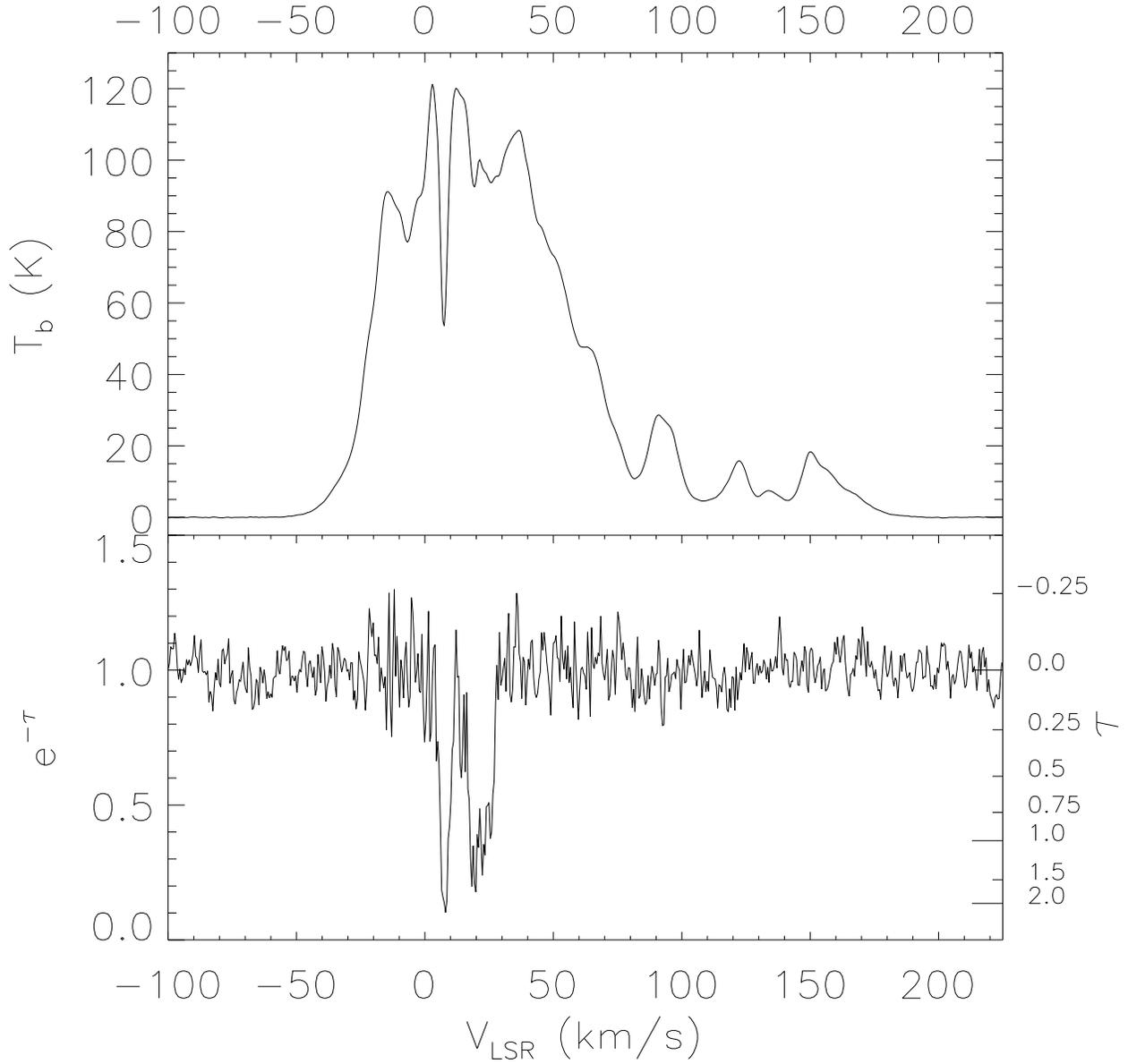}
\caption{\textit{Top}: The \ion{H}{1} emission spectrum (pulsar
``off'' spectrum) toward XTE~J1810--197.  \textit{Bottom}:  The
\ion{H}{1} absorption spectrum against XTE~J1810--197.  \label{fig:hiabs}
}
\end{figure}

The \ion{H}{1} absorption spectrum toward \magnetar\ is shown in
Figure~\ref{fig:hiabs}.  \magnetar\ is highly linearly polarized
\citep{xtepol} so that the YY polarization signal had a much better
signal-to-noise ratio (by a factor of $\approx 4$) than the XX
polarization.  The two polarizations were averaged together with weights
given by their signal-to-noise ratios to produce the spectrum shown.
The spectrum has not been smoothed in any way and shows the native
resolution of the observations.

\begin{figure}
\includegraphics[angle=0, width=6in]{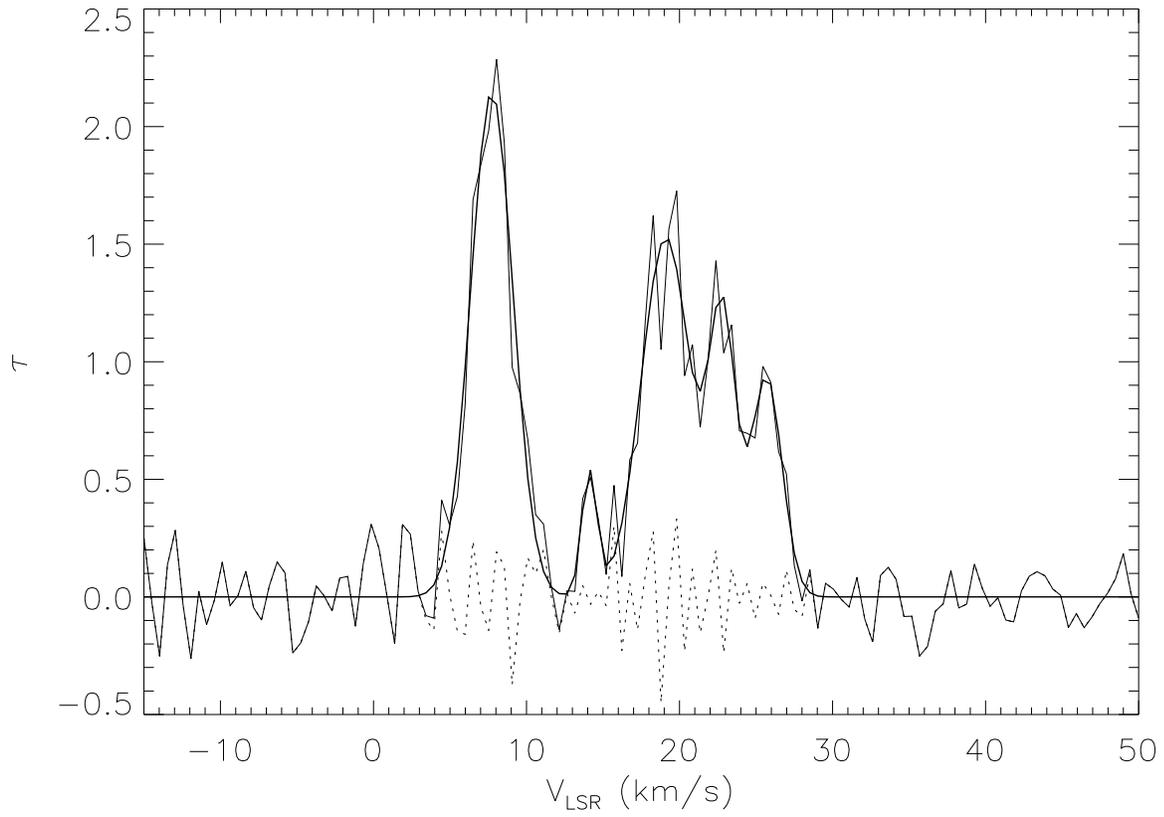}
\caption{Gaussian fits (\textit{thick line}) to the \ion{H}{1} opacity
(\textit{thin line}) observed toward XTE~J1810--197.  Residuals to the
fits are shown as the dotted line.  Results of the fits are listed in
Table~\ref{table:higaussians}.  \label{fig:gaussians} }
\end{figure}

Five Gaussians were fitted to the opacities determined from the \ion{H}{1}
absorption toward \magnetar.  The results of the fit are shown in
Table~\ref{table:higaussians} and in Figure~\ref{fig:gaussians}.
The columns in Table~\ref{table:higaussians} give for each line,
respectively, the opacity, Local Standard of Rest velocity ($V_{\rm
LSR}$), and full-width at half-maximum (FWHM).  The errors listed in
Table~\ref{table:higaussians} are $1\,\sigma$ errors from the Gaussian
fits.

The $V_{\rm LSR} = 7.7$\,km\,s$^{-1}$ absorption line is associated
with the Heeschen Cloud \citep{riegel}, which has also been called the
Riegel--Crutcher Cloud by \citet{rccloud}. The Heeschen Cloud is a nearby,
$d \approx 125\pm25$\,pc, cold cloud ($T_{\rm spin} \approx 40$\,K)
that covers Galactic longitudes $345^\circ$ to $25^\circ$ and latitudes
$\pm6^\circ$.  This cloud is also seen as a self-absorption feature in
the \ion{H}{1} emission spectrum (top panel of Fig.~\ref{fig:hiabs}).
The weak absorption line at $V_{\rm LSR} = 14.1$\,km\,s$^{-1}$ can be
kinematically associated with gas in the Carina--Sagittarius spiral arm
assuming that the line arises on the near side of the tangent point and
using the Galactic rotation model of \cite{englemaier}.  Likewise, the
$V_{\rm LSR} = 22.8$ and 25.7\,km\,s$^{-1}$ lines can be kinematically
associated with the Crux--Scutum spiral arm.

\begin{figure}
\plotone{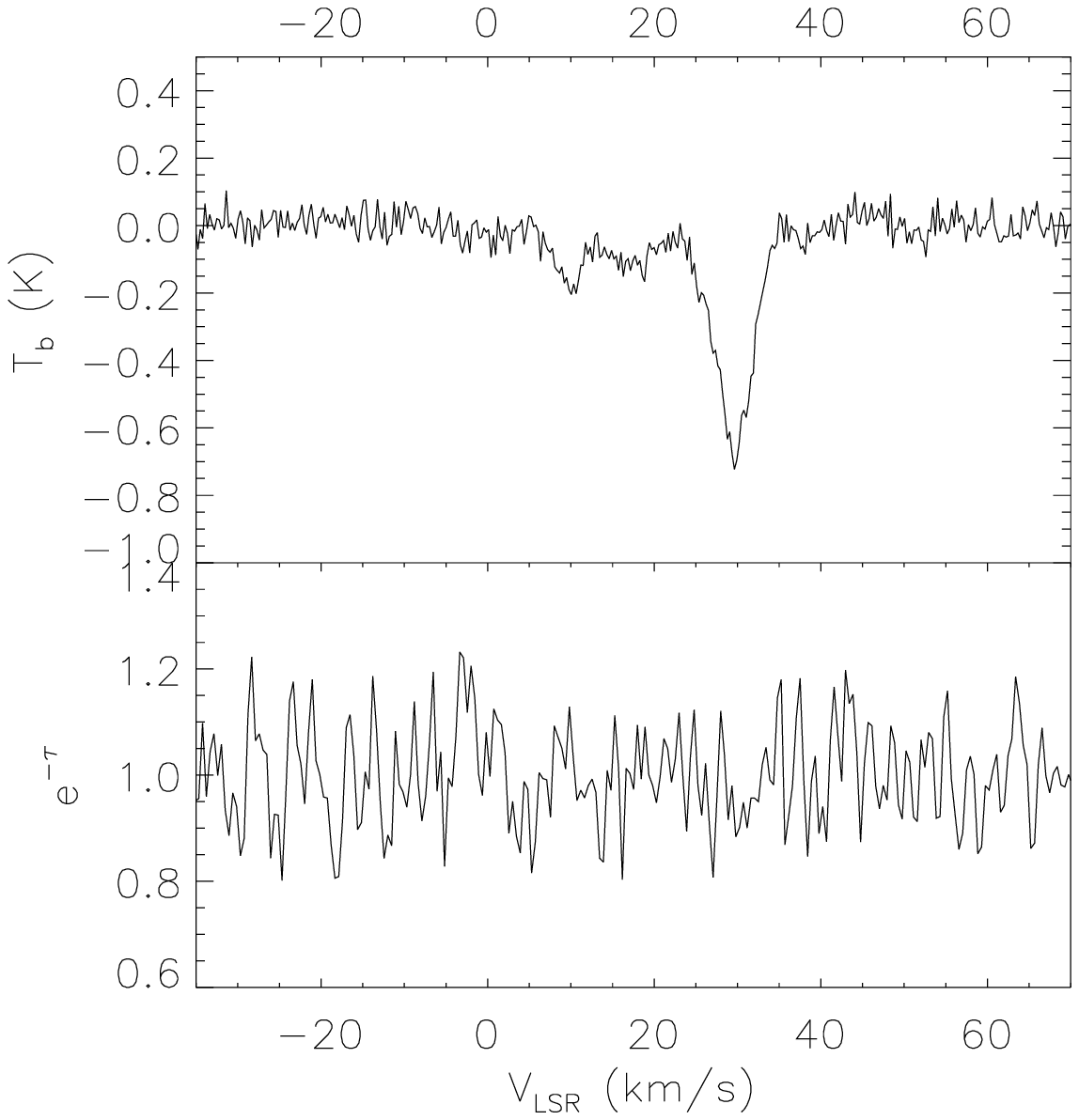}
\caption{\textit{Top}: The OH 1612\,MHz pulsar ``off'' spectrum
toward XTE~J1810--197.  \textit{Bottom}:  The OH 1612\,MHz
absorption spectrum against XTE~J1810--197.  \label{fig:oh1612} }
\end{figure}

\begin{figure}
\plotone{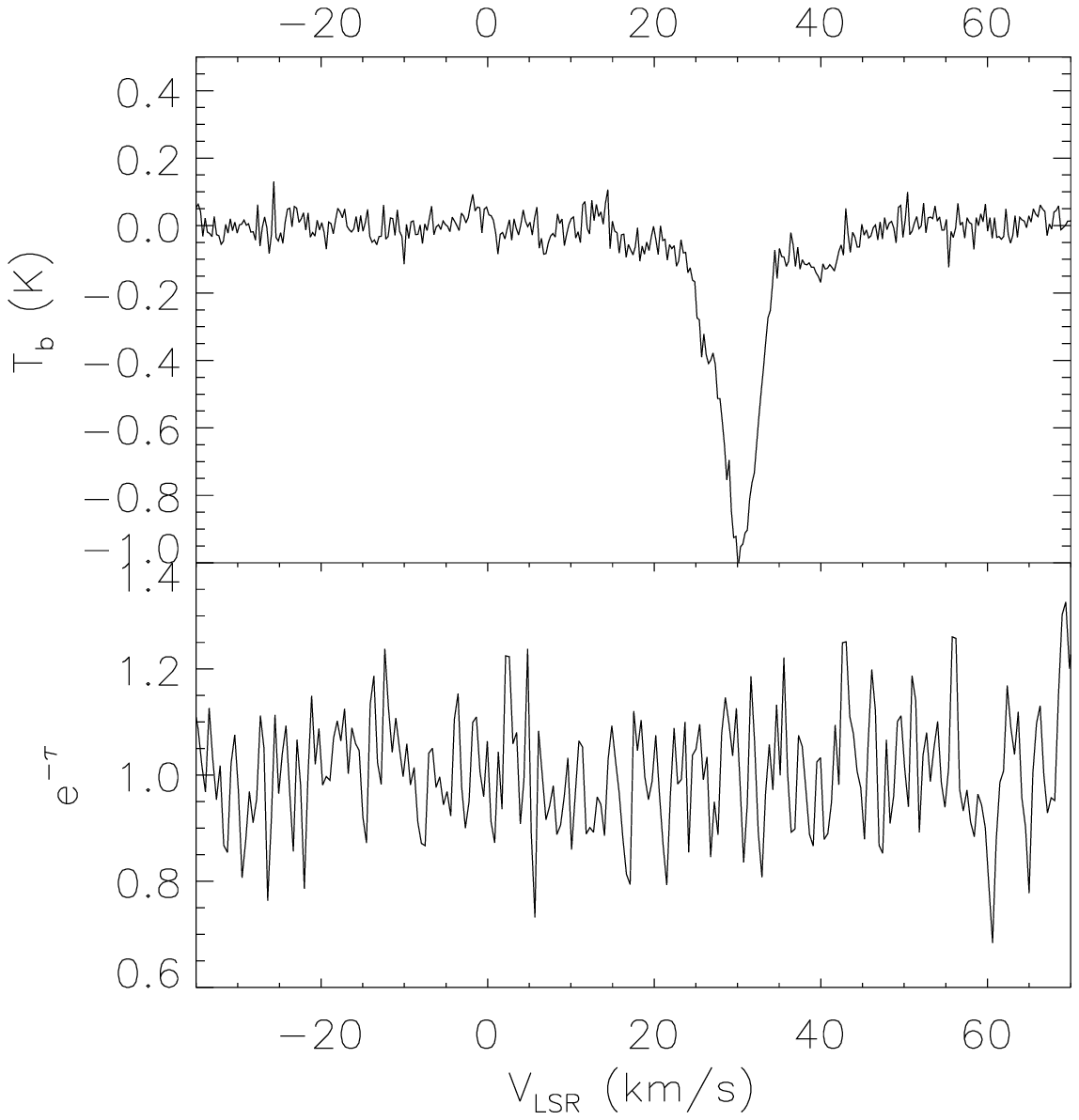}
\caption{\textit{Top}: The OH 1665\,MHz pulsar ``off'' spectrum
toward XTE~J1810--197.  \textit{Bottom}:  The OH 1665\,MHz
absorption spectrum against XTE~J1810--197.  \label{fig:oh1665} }
\end{figure}

\begin{figure}
\plotone{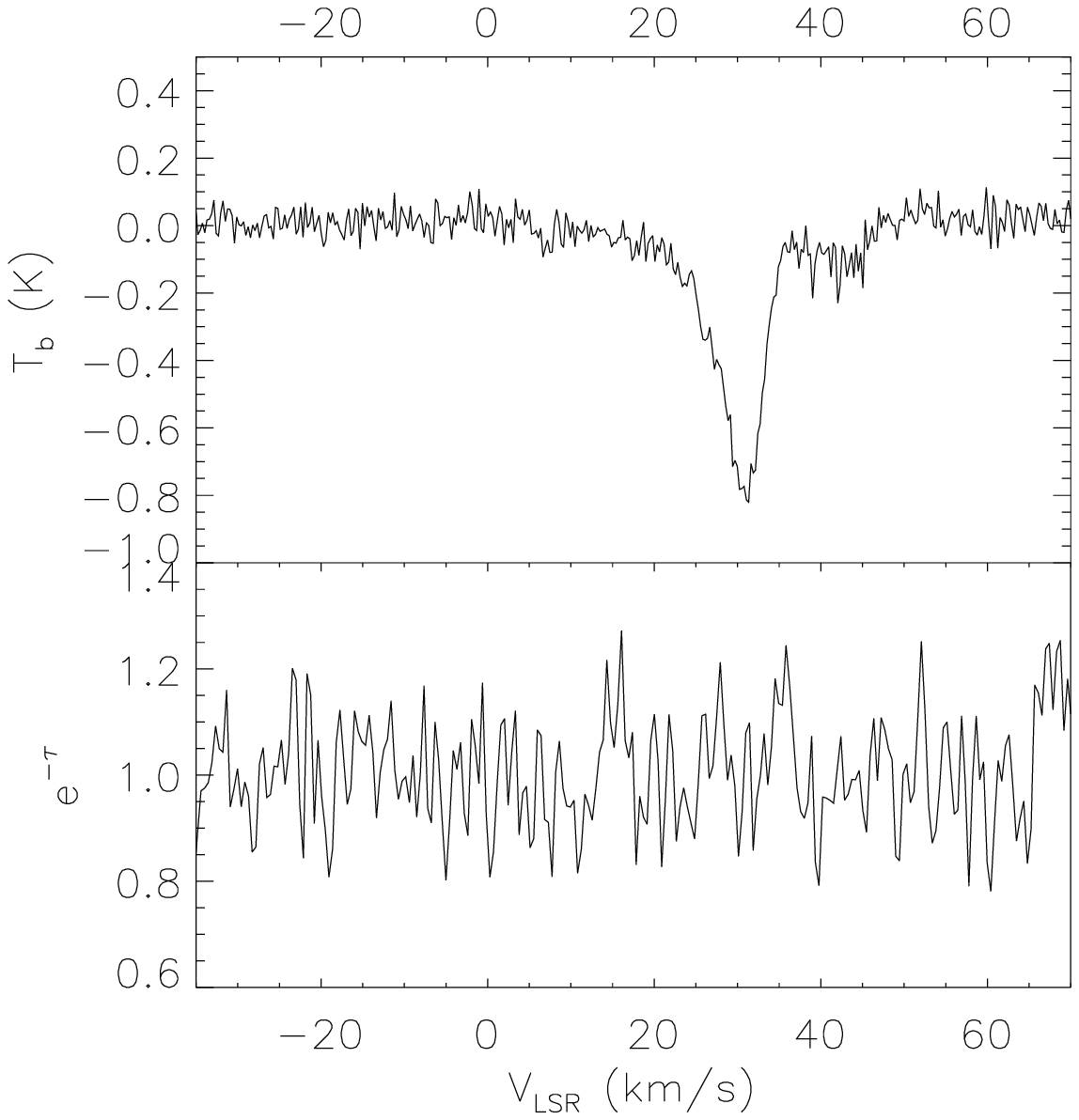}
\caption{\textit{Top}: The OH 1667\,MHz pulsar ``off'' spectrum
toward XTE~J1810--197.  \textit{Bottom}:  The OH 1667\,MHz
absorption spectrum against XTE~J1810--197.  \label{fig:oh1667} }
\end{figure}

\begin{figure}
\plotone{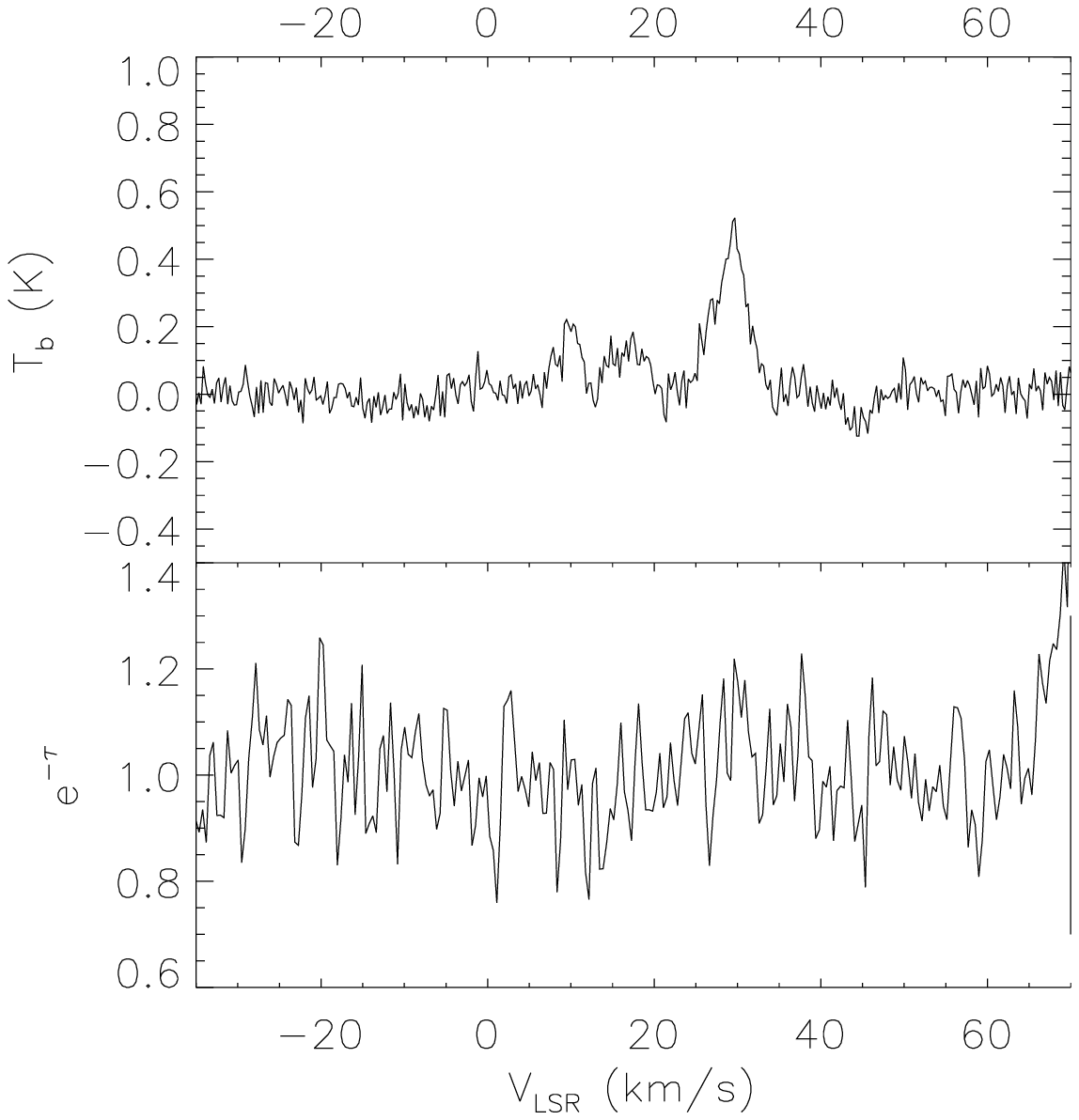}
\caption{\textit{Top}: The OH 1720\,MHz pulsar ``off'' spectrum
toward XTE~J1810--197.  \textit{Bottom}:  The OH 1720\,MHz
absorption spectrum against XTE~J1810--197.  \label{fig:oh1720} }
\end{figure}

The OH absorption spectra toward \magnetar\ are shown in
Figures~\ref{fig:oh1612}--\ref{fig:oh1720}.  With the strength of the
pulsar emission during the OH measurement epochs being much weaker than
was the case for the \ion{H}{1} measurement epoch \citep{fluxvarying},
the strong linear polarization of \magnetar\ means that only one
polarization effectively contributed to the measurement of the OH
absorption against \magnetar.  Thus, the OH absorption data shown
in Figures~\ref{fig:oh1612}--\ref{fig:oh1720} and discussed in this
paper are only from a single linear polarization (YY).  The $1\sigma$
opacity limits for any OH absorption toward \magnetar\ are given in
Table~\ref{table:ohlimits}.

\section{Determining the Kinematic Distance to XTE~J1810--197}

\subsection{Near or Far Side of the Velocity--Distance Relationship?}

\magnetar\ is located at Galactic coordinates $(l,b) = 10\fdg726,
-0\fdg158$.  At this Galactic longitude the kinematic velocity--distance
relationship is double-valued (see Fig.~\ref{fig:flat}).  Our first
concern is whether it is possible to determine on which side of the
tangent point \magnetar\ lies.  At this Galactic longitude, the tangent
point is about 8.3\,kpc from the Sun and has a $V_{\rm LSR} \approx
168$\,km\,s$^{-1}$ using the flat rotation curve of \cite{butler}.
Distance estimates for \magnetar\ based on its X-ray-fitted $N_{\rm H}$
are 2.5--5\,kpc \citep{firstdist, gotthelf, redclump}, and the DM-based
distance is 3.3\,kpc \citep{camilo}.  These strongly suggest that
\magnetar\ lies on the near side of the tangent point.

Another magnetar, \sgr, which lies $41'$ from \magnetar, has \ion{H}{1}
absorption at velocities greater than $V_{\rm LSR} = 50$\,km\,s$^{-1}$
and is thought to lie on the far side of the tangent point \citep{hiabs}.
Both \sgr\ and the Galactic SNR \snr\ \citep{becker} also have \ion{H}{1}
absorption features at negative velocities that must arise from the
far side of the Galactic disk if the rotation curve is flat.  However,
as the Galactic rotation model of \cite{sellwood} shows, these negative
velocities can also arise in the Galactic bar at a distance of 7\,kpc
toward these sources.  The Galactic bar induces large radial motions
that deviate from the normally assumed circular Galactic rotation.
Along the line of sight to \magnetar, the Galactic bar is responsible for
motions $> 90$\,km\,s$^{-1}$ more negative than the prediction of a flat
rotation curve (see Fig.~\ref{fig:weiner}).  Negative-velocity \ion{H}{1}
absorption is not found toward \magnetar, which according to this model
establishes an upper-limit to its distance of 7\,kpc (i.e., it is closer
than the Galactic bar).  This strengthens the conclusion that \magnetar\
lies on the near side of the tangent point.

\subsection{Is the Last Absorption Feature at the Actual Distance or a
Lower Limit?}

The number of \ion{H}{1} absorption features per kpc in the inner Galaxy
is about 1--3 \citep[][see their Table~3]{garwood}.  Assuming a flat
rotation curve and a distance of 5\,kpc, corresponding to a maximum
$V_{\rm LSR} \approx 40$\,km\,s$^{-1}$, along with a line width
of 3\,km\,s$^{-1}$ for the average \ion{H}{1} absorption feature,
we expect that 38\%--100\% of the velocities within $V_{\rm LSR} =
0$--40\,km\,s$^{-1}$ should contain \ion{H}{1} absorption along the line
of sight toward \magnetar.  If $\sim 38$\% of the velocity space were
occupied by absorption features, then the highest velocity feature in
the absorption spectrum toward \magnetar\ should be taken as indicating
a lower limit for the distance, since there is likely a significant
distance between \magnetar\ and the \ion{H}{1}-absorbing cloud nearest it.
If on the other hand the whole velocity range should contain \ion{H}{1}
absorption, it is likely that the last absorbing feature gives the actual
distance to \magnetar.

It is useful to look at absorption measurements toward objects near
\magnetar\ in order to help us determine the \ion{H}{1} absorption
velocity coverage.  The line of sight to \magnetar\ lies near that to
the giant Galactic \ion{H}{2} complex \wthirtyone\ \citep[see Fig.~1
of][]{corbel}.  \wthirtyone\ is comprised of several individual \ion{H}{2}
regions including \gtentwo, \gtenthree, and \gtensix.  Roughly on
the opposite side of \wthirtyone\ from \magnetar\ lies \gtenzero, the
wind-blown bubble of \lbv, and \sgr.  At higher Galactic longitude than
\magnetar\ lies the SNR \snr.  All of these objects have \ion{H}{1}
absorption measurements, except for \gtenzero\ and \lbv, which have
${\rm NH_3}$ absorption measurements \citep[][]{corbel}.

\citet[][]{becker} claimed that SNR~\snr\ lies well on the far side of the
tangent point.  However, \citet[][]{green} noted the lack of absorption
between $V_{\rm LSR} = 45$\,km\,s$^{-1}$ and the tangent point velocity,
which led them to assign a distance of $\approx 5$\,kpc to \snr, while
attributing weak absorption at negative velocities to peculiar motions
in local gas.  A distance of $\approx 5$\,kpc is also in reasonable
agreement with the expected expansion of the SNR \citep{green}.  If we
take the Galactic bar into account using the model of \citet{sellwood},
we can reconcile the \ion{H}{1} absorption with the expansion/age-derived
distance estimate of \citet{green}.  The \ion{H}{1} absorption to \snr\
at $V_{\rm LSR} \approx -20$\,km\,s$^{-1}$ arises from the Galactic bar,
which is at a distance of $\approx 5.5$--7.0\,kpc along this line of sight
(see Fig.~\ref{fig:weiner}).  With both the Galactic bar and \snr\ being
on the near side of the tangent point toward \snr, \ion{H}{1} absorption
is not expected at $V_{\rm LSR} > 45$\,km\,s$^{-1}$.  So we judge that
\snr\ likely lies within the Galactic bar at $\approx 5.5$--7\,kpc.

\citet[][]{corbel} determined that \gtenthree, \gtenzero, \lbv, and
\sgr\ all lie on the far side of the tangent point, while \gtentwo\ and
\gtensix\ lie on the near side.  The \ion{H}{1} absorption of \gtenthree\
\citep[Fig.~2 of][]{kalberla} and \sgr\ \citep[Fig.~2 of][]{cameron}
fill the range $V_{\rm LSR} = 0$--40\,km\,s$^{-1}$.  \ion{H}{1} toward
\gtensix\ shows absorption at all velocities within $V_{\rm LSR} =
0$--45\,km\,s$^{-1}$ \citep[Fig.~32 of][]{caswell}.  Likewise, the
\ion{H}{1} absorption toward \gtentwo\ \citep[Fig.~4 of][]{greisen} and
SNR \snr\ \citep[Fig.~3 of][]{becker} completely cover the velocity range
$V_{\rm LSR} = 0$--45\,km\,s$^{-1}$.  Since \gtensix, \gtentwo\ and \snr\
all lie on the near side of the tangent point and are only $16\arcmin$,
$33\arcmin$ and $30\arcmin$, respectively, in projection from \magnetar,
we argue that all velocities between the Sun and \magnetar\ should show
\ion{H}{1} absorption.  Therefore, the highest-velocity absorption feature
seen along the line of sight toward \magnetar\ gives us its true distance,
rather than a lower limit.

\subsection{The Kinematic Distance to XTE~J1810--197 }

For the kinematic distance models that we discuss we will assume that
the Sun is located 8.5\,kpc from the Galactic center.  We will also use
a velocity of $V_\sun=220$\,km\,s$^{-1}$ as the azimuthal velocity of
the LSR about the Galactic center.  These are the current IAU standards.
However, there is evidence that these values may not be correct \citep[see
the discussions in][]{gcdist,englemaier}, which would require a scaling
of the kinematic distance estimates presented below.

In Figure~\ref{fig:flat} we show the determination of the distance
to \magnetar\ using the flat rotation curve of \citet[][]{butler}.
Cold \ion{H}{1} clouds have random motions superposed on the uniform
Galactic rotation as indicated by their measured velocity dispersion
of 7\,km\,s$^{-1}$ \citep{jay}.  The dotted lines on either side of
the rotation model in Figure~\ref{fig:flat} indicate the deviation from
Galactic rotation of $\pm 7$\,km\,s$^{-1}$.  (The random motions of the
clouds are relative to the Galactic rotation and not to the measured
velocity of the cloud.  Thus, the 7\,km\,s$^{-1}$ represents an error
in converting the model's distance into a velocity  and not in using
the measured velocity of the cloud to obtain a distance as is usually
presumed.  This distinction can be quite important in determining the
distance and its errors when the observed velocity is at or near a
local maximum or minimum in the model. However, in most cases it makes
only a minor difference.)  The arrow at a constant $V_{\rm LSR}$ in
Figure~\ref{fig:flat} indicates the velocity of the highest velocity
\ion{H}{1} absorption.  The shaded regions in Figure~\ref{fig:flat}
indicate the allowed kinematic distances from the flat rotation curve.
Figures~\ref{fig:weiner} and \ref{fig:englemaier} are similar to
Figure~\ref{fig:flat}, using instead the Galactic rotation models
of \citet[][]{sellwood} and \citet[][]{englemaier}\footnote{We have
scaled the \citet[][]{englemaier} model for $R_\sun=8.5$\,kpc.},
respectively.

For the flat rotation curve used in Figure~\ref{fig:flat}, the last
\ion{H}{1} absorption feature toward \magnetar\ at $V_{\rm LSR} =
25.7$\,km\,s$^{-1}$ (see Table~\ref{table:higaussians}) gives a kinematic
distance of \flatdist\ or $13.3_{-0.5}^{+0.7}$\,kpc.  We can rule out
the larger distance since this is on the far side of the tangent point.
Although a flat rotation model may provide a reasonable distance estimate
for \magnetar, there are Galactic structures along the line of sight that
also need to be taken into account.  Spiral arms can have a substantial
effect on the expected Galactic rotational velocities \citep[][]{roberts}.
The Galactic bar may also have an influence by adding non-circular
motions to the general Galactic rotation.

The rotation model of \cite{sellwood} used in Figure~\ref{fig:weiner}
takes into account the effects of a strong potential associated with the
Galactic bar.  This model uses a global Galactic gravitational potential
to determine the motions of the gas in the ISM.  It uses both the
minimum and maximum observed \ion{H}{1} velocities along lines of sight
in the inner Galaxy in fitting the properties of the Galactic potential.
We can again immediately rule out the distance on the far side of the
tangent point.  Since no \ion{H}{1} absorption is seen at velocities
between 30 and 40\,km\,s$^{-1}$ (see Table~\ref{table:higaussians}) we
can eliminate all distances except \weinerdist.  The \citet[][]{sellwood}
model, however, does not take into account the effects of spiral arms.
Since magnetars are expected to lie in or near spiral arms this could have
a significant impact on the velocity-determined distance to \magnetar.

The rotation model of \cite{englemaier} used in
Figure~\ref{fig:englemaier} takes into account the effects of a
potential associated with the Galactic bar as well as those associated
with spiral arms.  The \citet[][]{englemaier} model was determined
in a similar fashion as the \citet[][]{sellwood} model.  However, in
the \citet[][]{englemaier} model only the maximum (minimum) observed
velocities for CO were used for positive (negative) Galactic longitudes.
It should be noted that CO observations do not trace column densities as
low as \ion{H}{1} does.  Combined with fitting to only one side of the
velocity--longitude data, this means that the \citet[][]{englemaier}
model uses a weaker potential for the Galactic bar compared with the
\citet[][]{sellwood} model.  In fact, the \citet[][]{englemaier} model
predicts that there should be no negative velocity gas along the line
of sight toward \magnetar, while the \citet[][]{sellwood} model does.
As can be seen in the Southern Galactic Plane Survey \ion{H}{1} data
\citep{sgpshi}, there is plenty of gas at negative velocities along this
line of sight.  We suggest that the \citet[][]{englemaier} Galactic bar
potential is too weak and that the \citet[][]{sellwood} model should be
used for velocities and distances associated with the Galactic bar.

We can, however, still use the \citet[][]{englemaier} model to determine
a distance for \magnetar\ since it is likely located in front of the
Galactic bar as is evidenced by the \citet[][]{sellwood} model results
from above.  In fact, comparing the \citet[][]{englemaier} result with
the \citet[][]{sellwood} result provides an indication of how spiral arms
may be affecting the velocity--distance relationship toward \magnetar.
We can again rule out distances that lie on the far side of the tangent
point and distances that include 30--40\,km\,s$^{-1}$ gas along the line
of sight.  As can be seen from Figure~\ref{fig:englemaier}, we obtain
kinematic distances of \englemaierdistboth.  Due to uncertainties in the
modeling of the Galactic rotation curve, the $0.1$\,kpc gap in distance
between these two values is in effect negligible.  We thus combine these
two possible distance ranges into a single value, \englemaierdist.

All of the above Galactic rotation models are derived from empirical
models fit to observed data.  There is one model that is fully derived
from observational data, that of \citet[][]{brand}.  In this model,
observations of \ion{H}{2} regions are used.  The velocities of
the \ion{H}{2} regions are determined from recombination lines and
\ion{H}{1} absorption.  Independent measurements provide distances to the
\ion{H}{2} regions.  Using the rotation model of \citet{brand}, shown
in Figure~\ref{fig:brand}, we obtain $d = 2.4\pm0.5$\,kpc. However,
this model does not have enough \ion{H}{2} regions at $d \ga 2$\,kpc
in the general direction of \magnetar\ to be able to provide a good
velocity--distance relation \citep[see][]{brand}, so we give it little
weight.

We conclude that the best estimate of a kinematically determined
distance to \magnetar\ is provided by the \citet[][]{sellwood} and
\citet[][]{englemaier} models, giving $d=$ \ourdistance.

\section{The Neutral Hydrogen Toward XTE~J1810--197: GBT Versus SGPS}

The GBT has a resolution of $9\farcm2$ at 21\,cm while the Southern
Galactic Plane Survey \citep[SGPS,][]{sgpshi} has a resolution of about
$3\farcm3$.  In Figure~\ref{fig:gbtsgps} we compare the GBT \ion{H}{1}
spectrum with the SGPS spectrum.  For the \ion{H}{1} self-absorption
associated with the Heeschen Cloud we find the remarkable result that the
line width of the absorption increases with increasing spatial resolution!
To our knowledge this effect has never been observed before for \ion{H}{1}
on arc-minute resolutions.

Inspection of the SGPS \ion{H}{1} data cube shows that there is structure
within the \ion{H}{1} emission inside of the GBT beam.  We convolved
the SGPS data with a beam the size of GBT's, so that the two data
sets would have the same spatial resolution.  As can be seen from
Figure~\ref{fig:gbtsgps} the GBT spectrum and $9\farcm2$ resolution SGPS
spectrum are in very good agreement.  If we compare the line width of the
\ion{H}{1} absorption due to the Heeschen Cloud at $\sim 8$\,km\,s$^{-1}$
with the SGPS data having \ion{H}{1} absorption seen against \magnetar,
we find that the line widths are identical.

These properties suggest that there is definite spatial structure in
the \ion{H}{1} emission on size scales of $3\farcm3$ to $9\farcm2$
(0.12--0.33\,pc for a Heeschan Cloud distance of $d=125$\,pc).  
That the tiny beam of absorption
against the magnetar (limited by the size of the pulsar emission region,
which is smaller than the pulsar's light-cylinder radius, or about
1\,light-second) has the same line width as the $0.12$\,pc-wide beam
of the SGPS at the Heeschen Cloud, suggests that there are few if any
spatial structures left unresolved by the SGPS in the Heeschen Cloud.
Since the \ion{H}{1} self-absorption line width does change between the GBT
and SGPS resolutions we can infer that either the absorption feature
comprises multiple narrow line features or that the non-thermal broadening
of the absorption feature changes between different structures in the
Heeschen cloud.  Attempts to fit the different line-width absorption-line
structures in the Heeschen cloud with multiple Gaussian components
does not improve the fitting, which suggests that a single Gaussian
is sufficient.  This implies that the non-thermal contribution to the
line width varies within the cloud.

The standard assumption is that non-thermal line broadening arises
from turbulence.  If the turbulence is intermittent, we can expect that
the turbulence has decayed more in some places than in other locations
\citep{frisch}.  This can then easily explain the observed line widths of
the Heeschen Cloud \ion{H}{1} absorption.  The areas that have undergone
more damping of the turbulence will have less turbulent energy and thus
have narrower non-thermal line widths.

\section{OH Absorption Limits Toward XTE~J1810--197}

Although we did not detect any OH absorption against \magnetar, the limits
are still interesting.  All previous OH absorption detections against
pulsars \citep[][]{snezana,joel,toney} have found that the absorption is
deeper (larger opacity) and has a narrower line width than the pulsar
``off'' spectra: in the three known cases, the OH absorption against
the pulsar is 2--3 times deeper than seen in the pulsar off spectra.

For our OH absorption limits against \magnetar\ of $\tau < 0.1$
($1\,\sigma$), we would expect our pulsar off spectra to have OH opacities
$\la 0.033$.  In Table~\ref{table:ohofftau} we list the opacities of the
OH lines in the pulsar off spectra.  We subtracted the system temperature
from the raw OH spectra, and then fitted a third order polynomial to
determine the continuum emission levels.  Dividing the spectra by the
continuum emission results in $e^{-\tau}$ spectra for the pulsar off
spectra.

If we assume that the OH pulsar off spectra should have a factor of
2--3 weaker opacity than the OH absorption against the pulsar, then
we were likely within a factor of about 2 of detecting OH absorption
against \magnetar\ at velocities within the range observed for \ion{H}{1}
absorption (Table~\ref{table:higaussians}), e.g., at 9.9\,km\,s$^{-1}$
at 1612\,MHz, and 9.8 and 16.9\,km\,s$^{-1}$ at 1720\,MHz (see
Table~\ref{table:ohofftau}).  Unfortunately, due the decay of the flux
of \magnetar\ \citep{fluxvarying}, we were not able to detect any OH
absorption.

Our OH absorption limits toward \magnetar\ are also meaningful for the
OH seen at $V_{\rm LSR} \ga 28$\,km\,s$^{-1}$.  The 1665\,MHz OH feature
at 30.7\,km\,s$^{-1}$ should have been detected at 2--$3\,\sigma$ if
it were between us and \magnetar.  However, we would not have expected
to detect OH absorption at these velocities based on the \ion{H}{1}
absorption results (Table~\ref{table:higaussians}).

\section{The Column Density to XTE~J1810--197}

From the \ion{H}{1} and OH observations that we have performed, it
is possible to estimate the column density of hydrogen $N_{\rm H}$,
both atomic and molecular, to \magnetar.  This can then be compared
with the range of values $N_{\rm H} = (6.5$--14$)\times 10^{21}$\,cm$^{-2}$
determined from fitting X-ray spectra \citep[][]{gotthelf,redclump}.

\subsection{Estimate of the Atomic Column Density to XTE~J1810--197}

We cannot directly measure the \ion{H}{1} column density toward \magnetar,
since we do not have enough constraints to perform radiative transfer
modeling in this direction.  We can however still make an estimate of the
column density to \magnetar.  To do this, we just integrate the \ion{H}{1}
emission spectrum between 0 and 25.7\,km\,s$^{-1}$.  Assuming that half
of the emission at a given velocity
comes from \ion{H}{1} on the near side of the tangent
point and the other half comes from gas with similar properties on
the far side of the tangent point,
we obtain an estimate of the column density to \magnetar,
\begin{equation}
N_{\rm HI} = {1 \over 2} \int_{0}^{25.7} 1.83 \times 10^{18}~T_{\rm B}(v)~dv.
\end{equation}
Since there are clouds on the near side of the tangent point that
significantly absorb the \ion{H}{1} emission from the far side of the
tangent point, as is evidenced by the absorption seen against \magnetar,
this method provides a lower limit to the column density of \ion{H}{1}
to \magnetar.  Upon performing the integration we find that $N_{\rm HI}
\ga 1.8 \times 10^{21}$\,cm$^{-2}$.

\subsection{Estimate of the Molecular Column Density to XTE~J1810--197}

We can use the pulsar ``off'' OH spectra to make an estimate of the
molecular column density to \magnetar.  From Figures~\ref{fig:oh1612}
and~\ref{fig:oh1720} we see that the 1612\,MHz OH lines are in
absorption while the 1720\,MHz OH lines are in emission, with both having
approximately the same amplitude.  Such conjugate emission arises in
regions where the OH column density is in the range $10^{14} < N_{\rm OH}/
\Delta v < 10^{15}$ cm$^{-2}$\,km$^{-1}$\,s with $\Delta v$ the velocity
resolution of the observations \citep[][]{joel,ohbook}.  The total column
density of OH is found by integrating over the whole line.

Integrating over the OH spectrum between 0 and 25.7\,km\,s$^{-1}$ and
assuming that half of the OH emission is from beyond the tangent point,
we find
\begin{equation} 
(2.4\pm 0.3) \times 10^{14} < N_{\rm OH} < (2.4\pm 0.3)
\times 10^{15}~{\rm cm}^{-2}. \nonumber
\end{equation} 
Using standard abundances, the ratio of the number of OH molecules to
the number of hydrogen atoms is $N_{\rm OH}/N_{\rm H} = 6 \times 10^{-8}$
\citep[][]{ohbook}, which gives
\begin{equation}
(4\pm 0.7) \times 10^{21} < N_{\rm H} < (40\pm 7) \times 10^{21}~{\rm cm}^{-2} \nonumber
\label{eq:nhoh}
\end{equation}
between us and the magnetar. 

Measurements of the $^{12}$CO$(2\rightarrow 1)$ spectrum toward
\magnetar\ are available from \citet[][]{cosurvey}.  The total column
density of hydrogen in molecular form can be determined from the
$^{12}$CO$(2\rightarrow 1)$ spectrum using
\begin{equation}
{N_{\rm H} = 2 N_{\rm H_2} = X_{\rm CO} \int_0^{25.7} T_{\rm B}^{\rm CO}(v) dv}
\end{equation}
where $X_{{\rm CO}}$ is a conversion factor.  The commonly used
``standard'' value for this is $X_{\rm CO} = 2 \times 10^{20}$
cm$^{-2}$\,K$^{-1}$\,km$^{-1}$\,s.  However, $X_{\rm CO}$ is
known to vary depending on the line of sight \citep[][]{xfactor}.
The $^{12}$CO$(2\rightarrow 1)$ spectrum nearest to the line of sight
toward \magnetar\ gives $\int_0^{25.7} T_{\rm B}^{\rm CO}(v) dv = 28.75$
K\,km\,s$^{-1}$, of which we will assume that half arises from beyond the
tangent point.  From Figure~4 of \citet[][]{xfactor} we see that $X_{\rm
CO} = 10^{20}$ cm$^{-2}$\,K$^{-1}$\,km$^{-1}$\,s is reasonable for the
amount of $^{12}$CO$(2\rightarrow 1)$ toward \magnetar.  This then yields
$N_{\rm H} \ga 2.8 \times 10^{21}$\,cm$^{-2}$.

This limit for $N_{\rm H}$ obtained from the CO spectrum is consistent
with the range of values determined using the OH spectra
(eq.~\ref{eq:nhoh}).  This also tells us that the OH column
densities are near the lower limit of $N_{\rm OH} / \Delta v =
10^{14}$ cm$^{-2}$\,km$^{-1}$\,s.  Since the CO spectra have a higher
signal-to-noise ratio, we will use the CO value for the molecular gas
contribution to $N_{\rm H}$.

Adding the molecular and atomic components of $N_{\rm H}$, we obtain
$N_{\rm H} \ga 4.6 \times 10^{21}$\,cm$^{-2}$.  This limit is in
agreement with the values determined from the X-ray spectrum of \magnetar.
Unfortunately we only have a lower limit for the total column density
that is below the values determined from the X-ray spectrum.

\section{Models of the Line of Sight Toward XTE~J1810--197}

A detailed molecular model of the ISM toward \sgr\ has been developed
by \citet[][]{corbelone} and \citet[][]{corbel}, depicted in Figure~8 of
the latter.  With increasing distance toward \sgr, their model encounters
gas at $V_{\rm LSR}=4$, 24, 30, 38, 44, and then 13\,km\,s$^{-1}$ in
reaching the Scutum-Crux spiral arm \citep[labeled as the 30\,km\,s$^{-1}$
spiral arm in Fig.~8 of][]{corbel}.  Since the line of sight toward \sgr\
is very close to that of \magnetar\ we might expect to encounter clouds
at roughly the same velocities on the line of sight toward \magnetar.

The gas at 4\,km\,s$^{-1}$ toward \sgr\ is likely one of the two velocity
components of the Heeschen Cloud \citep[][]{riegel} and can be associated
with the \ion{H}{1} absorption feature at 7.7\,km\,s$^{-1}$ toward
\magnetar, which is from the other velocity component of the Heeschen
Cloud.  The gas at 24\,km\,s$^{-1}$ toward \sgr\ can be associated with
the \ion{H}{1} absorption features at 22.8 or 25.7\,km\,s$^{-1}$ toward
\magnetar.  We can associate the 13\,km\,s$^{-1}$ gas toward \sgr\ with
the \ion{H}{1} absorption seen at 14.1\,km\,s$^{-1}$ toward \magnetar.

If the model of \citet[][]{corbelone} and \citet[][]{corbel} is correct,
then we should also expect see \ion{H}{1} absorption at velocities of
roughly 30, 38, and 44\,km\,s$^{-1}$ toward \magnetar, since we observe
absorption that can be associated with the 13\,km\,s$^{-1}$ cloud in
their model.  In fact, we do not observe any \ion{H}{1} absorption
against \magnetar\ at these velocities.  This suggests that either
(1) the model of \citet[][]{corbelone} and \citet[][]{corbel} is not
entirely correct, or (2) it cannot be applied to the line of sight toward
\magnetar, which is only $41'$ away, or (3) there is molecular material
without \ion{H}{1} along these lines of sight.  The last possibility
is not likely since molecular clouds are expected to have cosmic-ray
ionization and photo-dissociation regions within and at their outer edges,
which would produce atomic hydrogen \citep[see][\S~9.2, and references
therein]{atomiccloud}.

\section{Discussion}

Using the $\mbox{DM} = 178\pm5$\,cm$^{-3}$\,pc measured for \magnetar\
\citep{camilo}, its distance according to the \citet{ne2001} electron
density model is 3.3\,kpc.  This model has a claimed average uncertainty
of about 20\% which, however, can be much larger for individual objects.
For the sake of discussion, we assume an uncertainty of 1\,kpc.
The electron density model was derived using distances to pulsars that
in many cases were determined via \ion{H}{1} absorption spectra assuming
a flat rotation curve, so that it seems most appropriate to compare the
DM--derived distance of $3.3\pm1$\,kpc with our flat rotation curve's
kinematic distance of \flatdist.  These two values agree remarkably well
and imply that along the line of sight to \magnetar, the \citet{ne2001}
model gives a good representation of the average free electron density
out to about 4\,kpc.

The distance to \magnetar\ has been estimated from X-ray observations
to range over 2.5--5\,kpc \citep{gotthelf,firstdist}.  These distances
are determined by converting $N_{\rm H}$ values obtained from fits to
the X-ray spectra into visual extinction, $A_V$, and an estimate of
the $A_V$ per kpc in the Galaxy.  This method is limited by a number of
complications: (1) the spectral model used to fit the X-ray data, e.g.,
two blackbodies versus a blackbody and a power law; (2) the $N_{\rm H}$
versus $A_V$ relationship determined locally (within $\sim 1$\,kpc) 
but used for large distances ($> 1$\,kpc);
(3) the large deviations from the fitted  $N_{\rm H}$
versus $A_V$ relationship for any particular line of sight;
and (4) the $d$--$A_V$ relationship also determined locally but used
for large distances.

\cite{redclump} determined $d = 3.1\pm0.5$\,kpc toward \magnetar\ assuming
an X-ray $N_{\rm H} = 14 \times 10^{21}$\,cm$^{-2}$, which is higher
than any of the values fitted by \cite{firstdist} or \cite{gotthelf}.
We consider the $d$--$A_V$ relation of \citet{redclump} using red clump
stars in the line of sight to \magnetar\ to be 
an improvement,
although it is still necessary to apply an X-ray--fitted value of $N_{\rm
H}$ to this relation and then convert it into a visual extinction
$A_V$.  Using $N_{\rm H} = 6.5 \times 10^{21}$\,cm$^{-2}$ determined
by \citet{gotthelf}, which is arguably a lower limit, the extinction
toward \magnetar\ becomes $A_V \approx$\,3--4.5\,mag using
Figure~3 of \citet{predehl}.  This then yields
a distance estimate of 2--3.5\,kpc based on Figure~7 of \cite{redclump}.

The kinematic distances determined from the \ion{H}{1} absorption
measurements presented in this paper rely only on the model of Galactic
rotation used to convert the measured velocity into a distance.
The \citet{sellwood} and \citet{englemaier} models both give consistent
results (see Table~\ref{table:distance}).  We prefer the kinematic
distance of \ourdistance\ determined in this paper over the DM- and
X-ray-derived distances, because our conversion of measured velocity
to distance is more direct and better constrained than through these
other methods.

In Figure~7 of \citet{redclump} we see that the extinction vs.\ distance
rises steeply between 3.0 and 3.5\,kpc and that for $A_V \le 13$ the
distance can be limited to be less than 4\,kpc.  The line of sight
toward \magnetar\ contains the Infrared Dark Cloud (IRDC) \darkcloud\
\citep[][]{irdc,carey}.  IRDCs are thought to be places where Giant
Molecular Clouds are either just starting to form massive stars or on the
verge of beginning to form stars.  The sudden increase in opacity vs.\
distance seen by \citet{redclump} is likely associated with this IRDC.
We note that the highest opacity regions of the IRDC cover less than
half the area on the sky that \citet{redclump} used to determine the
visual extinction vs.\ distance toward \magnetar.  Also, IRDCs can have
extremely high opacities such that only the edge of the cloud is seen
even in the infrared \citep{irdc,atomiccloud}.  Since the opacities can be
very large, it is plausible that \cite{redclump} may have underestimated
the actual extinction vs.\ distance toward \magnetar.

\cite{co} associated IRDC \darkcloud\ with $^{12}{\rm
CO}\left(1\rightarrow 2\right)$ emission centered at 32\,km\,s$^{-1}$,
implying $d<3.8_{-0.5}^{+0.5}$\,kpc for a flat rotation curve and
$d<4.4_{-0.3}^{+0.6}$\,kpc for the \cite{sellwood} and \cite{englemaier}
models.  Our OH spectra show a large spectral feature at the same
velocities that we also associate with the IRDC.  All observed \ion{H}{1}
absorption features lie at velocities lower than those associated with the
IRDC.  This puts \magnetar\ no further than the front edge of the IRDC.

We consider the 4\,kpc upper limit from Figure~7 of \citet{redclump}
to be a hard upper limit on the distance to \magnetar.  This further
constrains slightly the distance estimate of 3.1--4.3\,kpc obtained
from \ion{H}{1} absorption measurements (\S~4.3).  Overall, we can
thus summarize that the distance to \magnetar\ is $3.5\pm0.5$\,kpc.
Together with the measured proper motion of the AXP, this results in
a transverse velocity corrected to the LSR of $212\pm35$\,km\,s$^{-1}$
\citep{helfand}, a perfectly ordinary velocity among pulsars.

\acknowledgments

We would like to thank Jay Lockman for many useful discussions and
T. Dame who graciously provided us with the $^{12}$CO$(2\rightarrow 1)$
spectrum toward \magnetar.  The GBT is operated by the National Radio
Astronomy Observatory, a facility of the National Science Foundation
operated under cooperative agreement by Associated Universities, Inc.
F.C. acknowledges the NSF for support through grant AST-05-07376.

\clearpage

\begin{deluxetable}{lcc}
\tablewidth{0pt}
\tablecolumns{3}
\tablecaption{\label{table:absobs} Observations of \magnetar\ at GBT }
\tablehead{
\colhead{Date} &
\colhead{Time (UTC)} &
\colhead{Species} }
\startdata
2006 Jun 6 & 07:30 -- 11:00 & \ion{H}{1} \\
2006 Jul 22/23 & 23:00 -- 07:30 & OH \\
2006 Aug 31/Sep 1 & 23:30 -- 05:00 & OH \\
2006 Sep 2 & 00:00 -- 05:00 & OH \\ 
2006 Sep 4 & 00:00 -- 05:00 & OH \\ 
2006 Sep 24/25 & 19:45 -- 03:40 & OH \\ 
2006 Oct 21 & 17:00 -- 19:09 & OH \\
2006 Oct 21/22 & 20:55 -- 01:45 & OH
\enddata
\end{deluxetable}

\clearpage

\begin{deluxetable}{ccc}
\tablewidth{0pt}
\tablecolumns{3}
\tablecaption{\label{table:higaussians} Gaussian fits to the \ion{H}{1}
absorption lines toward XTE~J1810--197 }
\tablehead{
\colhead{$\tau$}         & 
\colhead{$V_{\rm LSR}$}  &
\colhead{FWHM}           \\
\colhead{}               &
\colhead{(km\,s$^{-1}$)} &
\colhead{(km\,s$^{-1}$)} }
\startdata
$2.15 \pm 0.09 $ & $  7.73 \pm 0.06 $ & $ 3.3 \pm 0.2 $ \\
$0.5  \pm 0.1  $ & $ 14.1  \pm 0.2  $ & $ 1.2 \pm 0.4 $ \\
$1.53 \pm 0.08 $ & $ 19.1  \pm 0.1  $ & $ 3.8 \pm 0.4 $ \\
$1.2 \pm 0.1 $ & $ 22.8  \pm 0.2  $ & $ 2.1 \pm 0.4 $ \\
$0.94 \pm 0.1 $ & $ 25.7  \pm 0.2  $ & $ 2.4 \pm 0.5 $
\enddata
\tablecomments{ See Figs.~\ref{fig:hiabs} and \ref{fig:gaussians} for
data upon which these fits are based. }
\end{deluxetable}

\clearpage

\begin{deluxetable}{cc}
\tablewidth{0pt}
\tablecolumns{2}
\tablecaption{\label{table:ohlimits} OH absorption limits toward 
XTE~J1810--197 }
\tablehead{
\colhead{$\nu$} &
\colhead{$\sigma_\tau$\tablenotemark{a}} \\ 
\colhead{(MHz)} &
\colhead{} } 
\startdata
1612 & 0.09 \\
1665 & 0.10 \\
1667 & 0.10 \\
1720 & 0.10
\enddata
\tablenotetext{a}{$1\,\sigma$ limits.}
\end{deluxetable}

\clearpage

\begin{deluxetable}{cccc}
\tablewidth{0pt}
\tablecolumns{3}
\tablecaption{Gaussian fits to the OH opacity
in the pulsar ``off'' spectra toward XTE~J1810--197 \label{table:ohofftau}}
\tablehead{
\colhead{$\nu$} &
\colhead{$\tau$}         & 
\colhead{$V_{\rm LSR}$}  &
\colhead{FWHM}           \\
\colhead{(MHz)}    &
\colhead{}               &
\colhead{(km\,s$^{-1}$)} &
\colhead{(km\,s$^{-1}$)} }
\startdata
1612 & $0.026\pm 0.002$ & $9.9\pm 0.2$ & $4.4\pm 0.6$ \\
1612 & $0.097\pm 0.002$ & $29.54\pm 0.05$ & $5.4\pm 0.1$ \\
\hline
1665 & $0.059\pm 0.009$ & $28.7\pm 0.5$ & $7.9\pm 0.4$ \\
1665 & $0.10\pm  0.01$ & $30.69\pm 0.08$ & $3.9 \pm 0.3$ \\
1665 & $0.020\pm 0.002$ & $39.9\pm 0.3$ & $5.5\pm 0.8$ \\
\hline
1667 & $0.045\pm 0.005$ & $28.1\pm 0.6$ & $9.7\pm 0.6 $ \\
1667 & $0.086\pm 0.007$ & $31.11\pm 0.08 $ & $4.3\pm 0.3$ \\
1667 & $0.018\pm 0.002$ & $42.0\pm 0.4 $ & $6.3\pm 0.9$ \\
\hline
1720 & $-0.023\pm 0.002$ & $9.8\pm 0.1$ & $3.1\pm 0.3$ \\
1720 & $-0.020\pm 0.002$ & $16.9\pm 0.2$ & $4.8\pm 0.5$ \\
1720 & $-0.021\pm 0.004$ & $26.2\pm 0.2$ & $2.3\pm 0.5$ \\
1720 & $-0.063\pm 0.002$ & $29.5\pm 0.1$ & $3.8\pm 0.3$ \\
1720 & $0.014\pm 0.002$ & $44.4\pm 0.3$ & $3.0\pm 0.6$
\enddata
\end{deluxetable}

\clearpage

\begin{deluxetable}{llrc}
\tabletypesize{\footnotesize}
\tablewidth{0pt}
\tablecolumns{4}
\tablecaption{\label{table:distance} Distance estimates for XTE~J1810--197 }
\tablehead{
\colhead{Measurement} & 
\colhead{Method}      &
\colhead{$d$}         &
\colhead{Refs.}       \\
\colhead{}            &
\colhead{}            &
\colhead{(kpc)}       &
\colhead{}            }
\startdata
DM & $n_e$ model \citep{ne2001} & $3.3 \pm 1$ & 1 \\
X-ray: blackbody + power law &  ${N_{\rm H} \rightarrow A_V \rightarrow 1.5-2.0~{\rm mag\,kpc}^{-1} \rightarrow } d$ & $\sim 5.0$ & 2 \\
X-ray: two blackbodies & ${N_{\rm H} \rightarrow A_V \rightarrow 1.5-2.0{\rm ~mag\,kpc}^{-1} \rightarrow } d$  & $\sim 2.5$ & 3 \\
X-ray: blackbody + power law & ${N_{\rm H} \rightarrow A_V \rightarrow {\rm red~clump~stars} \rightarrow } d$  & $3.1\pm0.5$ &  4 \\
X-ray: two blackbodies& ${N_{\rm H} \rightarrow A_V \rightarrow {\rm red~clump~stars} \rightarrow }d$  & 2--3.5 & 5 \\
\ion{H}{1} absorption & Flat rotation curve \citep{butler} & \flatdist & 5 \\
\ion{H}{1} absorption & \citet[][]{sellwood} model & \weinerdist &5\\
\ion{H}{1} absorption & \citet[][]{englemaier} model & \englemaierdist &5\\
\ion{H}{1} absorption & \citet[][]{brand} model & $2.4\pm0.5$&5
\enddata
\tablerefs{(1) \citet{camilo}; (2) \citet{firstdist}; (3)
\citet{gotthelf}; (4) \citet{redclump}; (5) this work. }
\tablecomments{See \S~4 for a discussion of the various rotation curve
models that we fit to the \ion{H}{1} data.  The X-ray fits done by other
authors are discussed in \S~9.  Overall, our best distance determination
comes largely from the direct \ion{H}{1} measurements on \magnetar,
and is $3.5\pm0.5$\,kpc (see \S~9). }
\end{deluxetable}

\clearpage

\clearpage

\clearpage

\clearpage

\clearpage

\clearpage

\clearpage

\begin{figure}
\includegraphics[angle=90, width=6in]{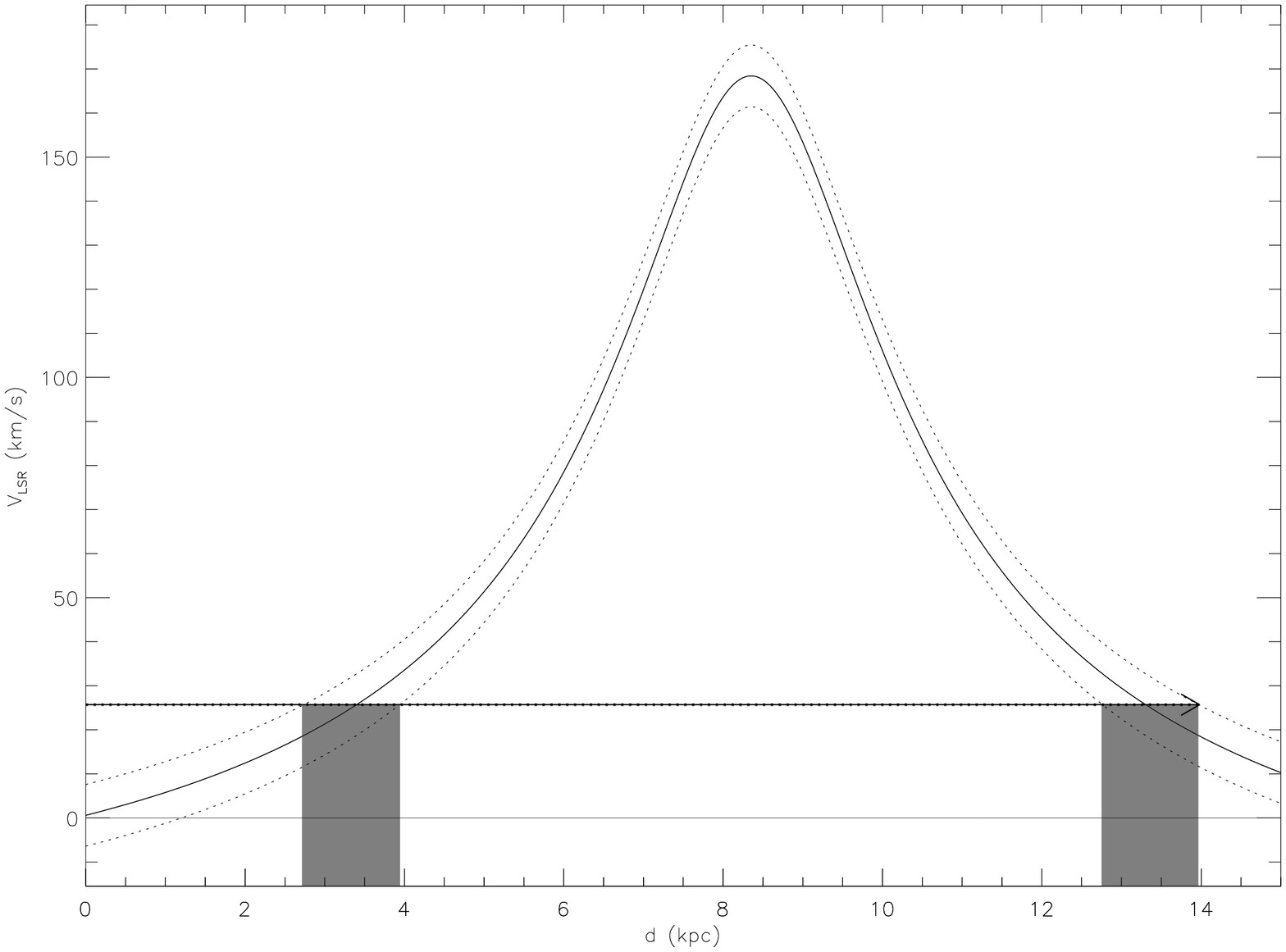}
\caption{The flat Galactic rotation model (\textit{solid curve}) of
\citet{butler}.  The $x$-axis is the distance from the Sun along the line
of sight of \magnetar.  The $y$-axis is the radial velocity of the gas.
The dotted lines indicate the deviation from Galactic rotation of $\pm
7$\,km\,s$^{-1}$.  The arrow at $V_{\rm LSR}=25.7$\,km\,s$^{-1}$ indicates
the highest velocity \ion{H}{1} absorption seen toward \magnetar\ (see
Table~\ref{table:higaussians}).  The shaded regions indicate the allowed
kinematic distances from the flat rotation curve (see discussion in
\S\S~4.2 and 4.3 for interpretation of these figures). \label{fig:flat} }
\end{figure}

\clearpage

\begin{figure}
\includegraphics[angle=90, width=6in]{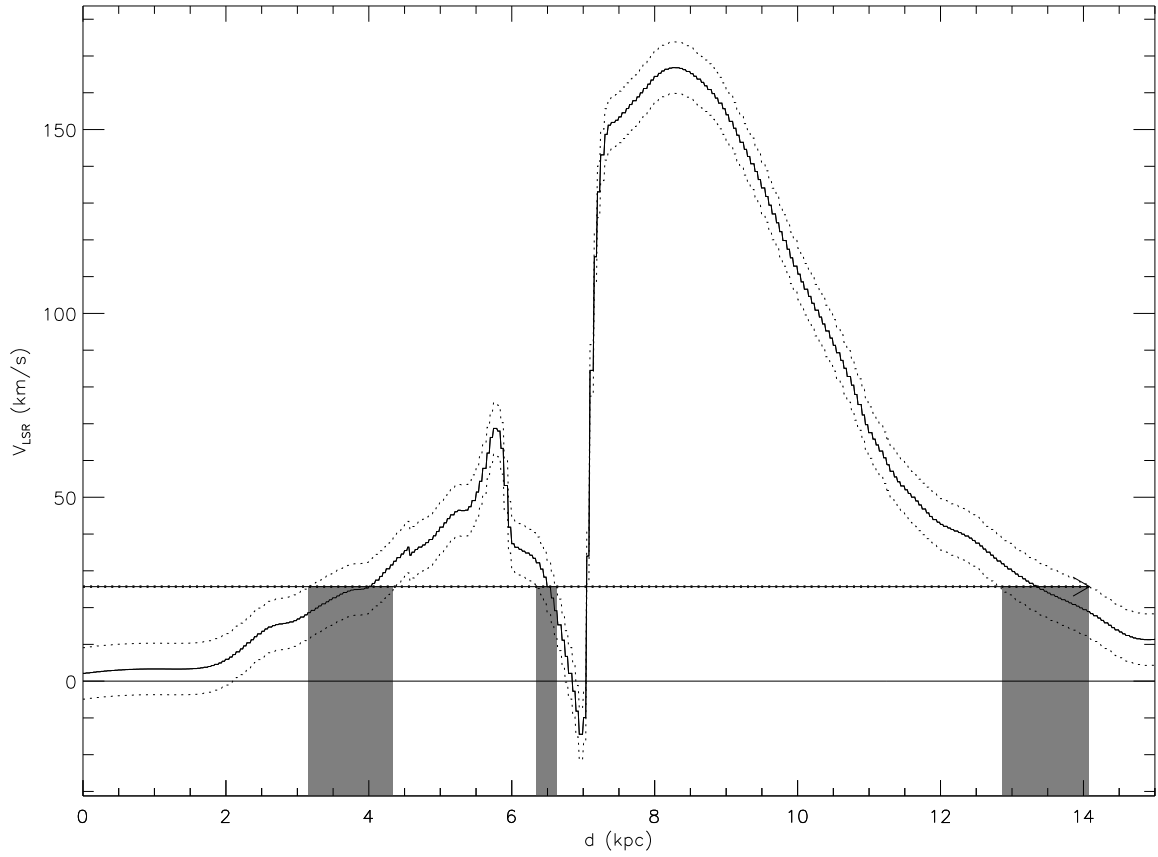}
\caption{The same as Fig.~\ref{fig:flat}, except that it uses the Galactic
rotation model of \citet[][]{sellwood}, considering potential due to a
bar. \label{fig:weiner} }
\end{figure}

\clearpage

\begin{figure}
\includegraphics[angle=90, width=6in]{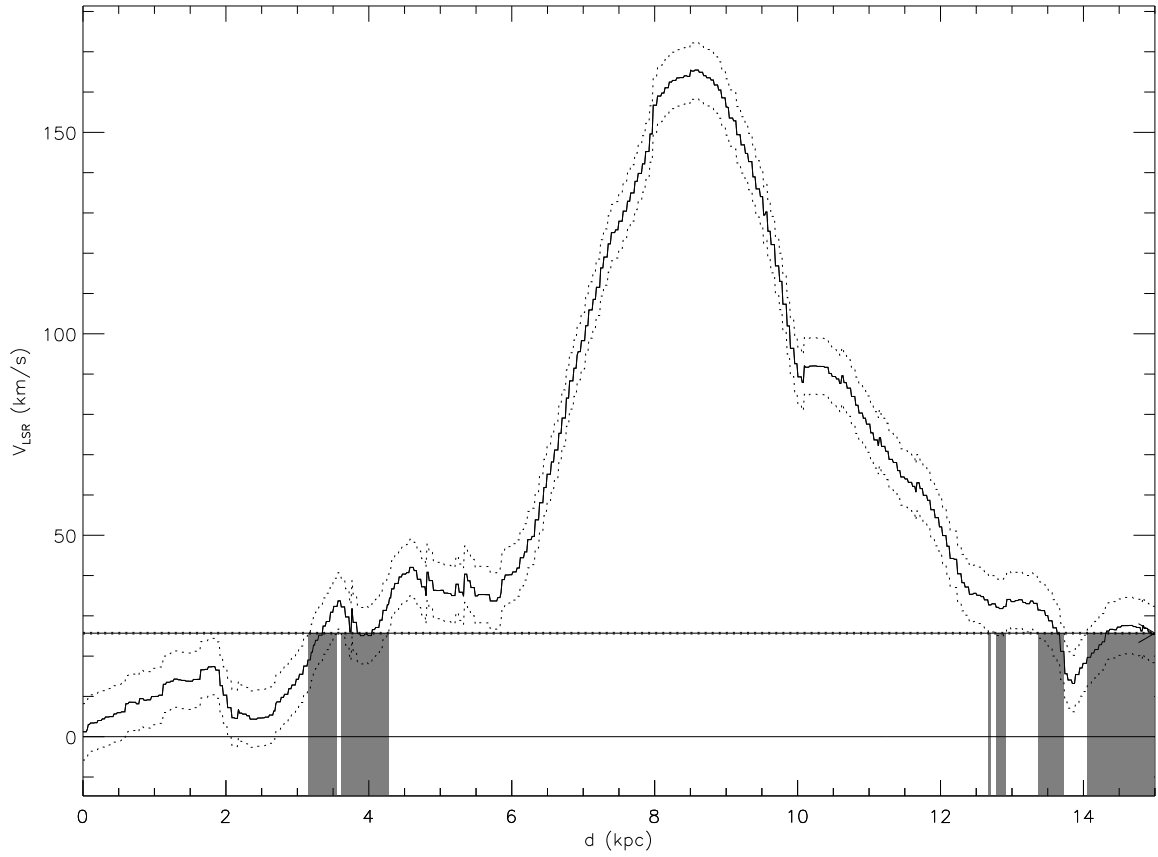}
\caption{The same as Fig.~\ref{fig:flat}, except that it uses the Galactic
rotation model of \citet[][]{englemaier}, considering potential due to
a bar and spiral arms. \label{fig:englemaier} }
\end{figure}

\clearpage

\begin{figure}
\includegraphics[angle=90, width=6in]{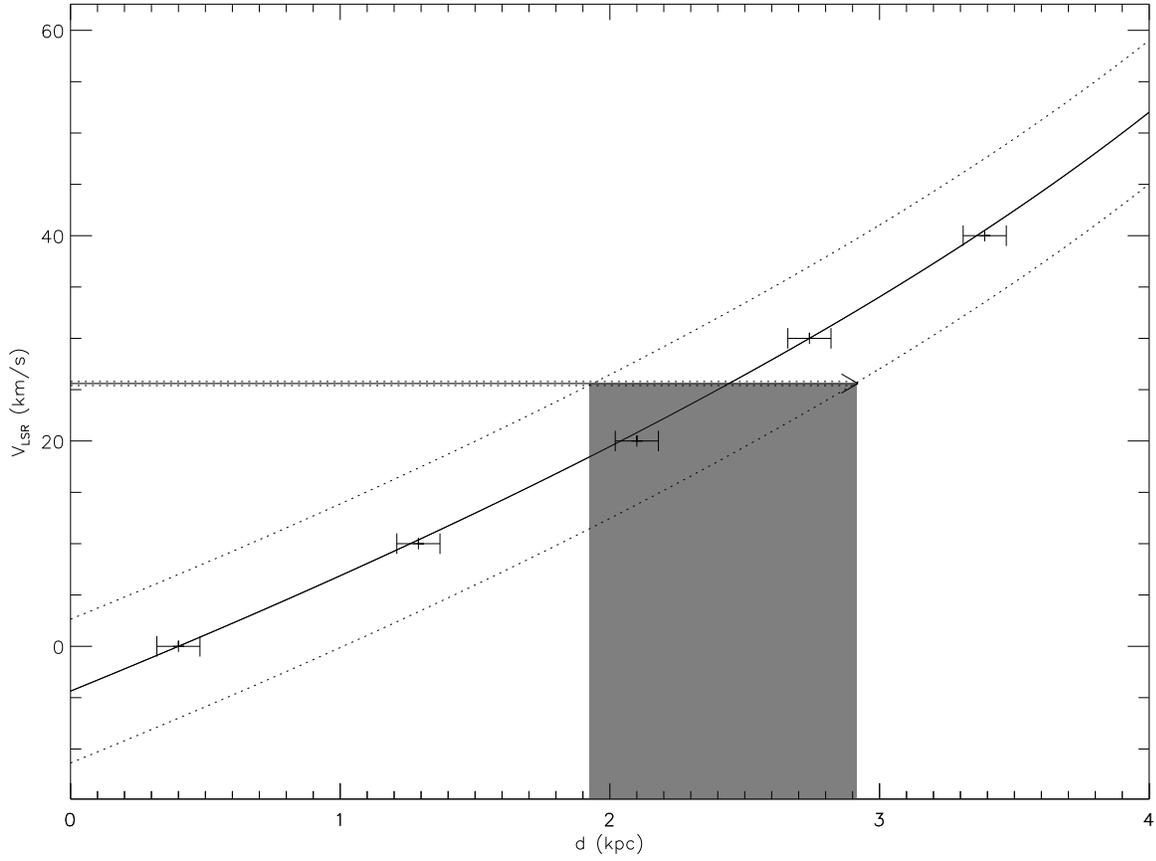}
\caption{The same as Fig.~\ref{fig:flat}, except that it uses the
Galactic rotation model of \citet{brand}, derived from observations
of \ion{H}{2} regions.  The points and error bars are derived from
Fig.~2 of \citet{brand}.  The solid line is a quadratic fit to the
points. \label{fig:brand} }
\end{figure}

\clearpage

\begin{figure}
\includegraphics[angle=0, width=6in]{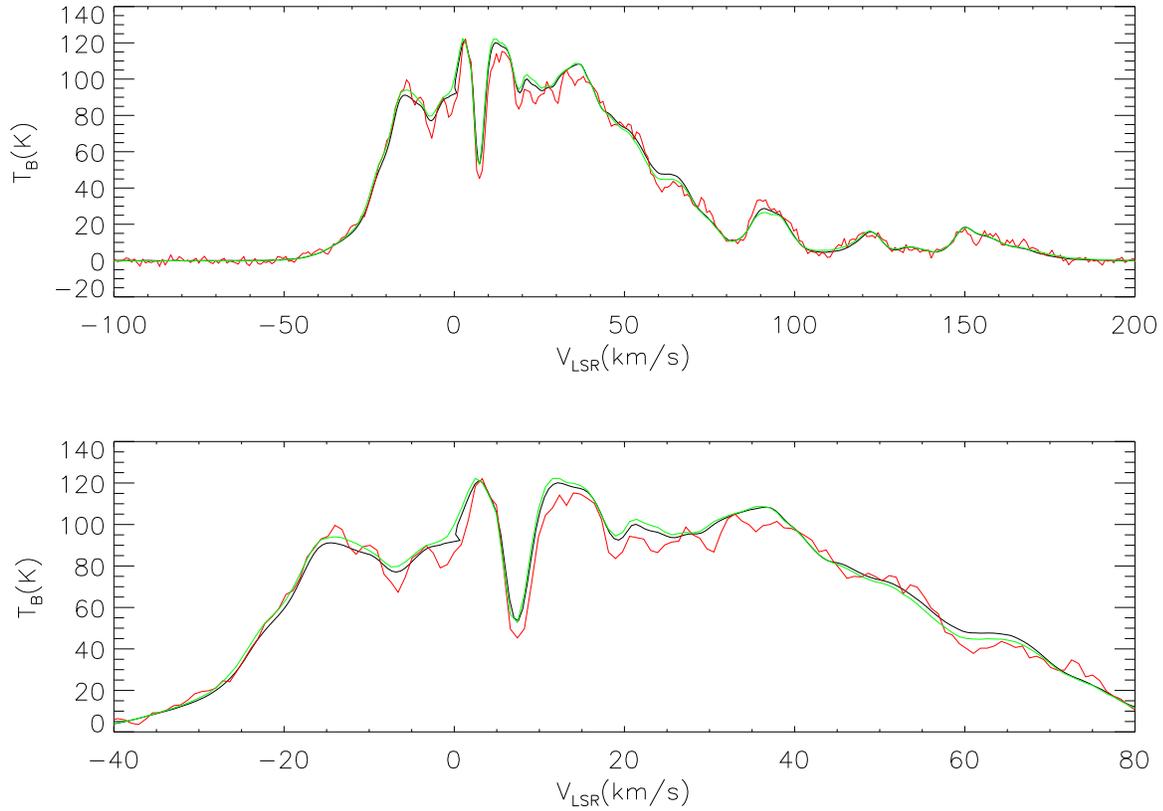}
\caption{Comparison of the GBT \ion{H}{1} spectrum (\textit{black line})
with the full-resolution SGPS \ion{H}{1} spectrum (\textit{red line})
toward \magnetar.  The SGPS spectrum convolved to the same resolution
as the GBT is shown as the green line.  The top panel shows the full
range of emission, while the lower panel is zoomed in to show the
difference in line widths for the Heeschen Cloud absorption around
8\,km\,s$^{-1}$. \label{fig:gbtsgps} }
\end{figure}

\end{document}